\documentclass[11pt]{article} 

\usepackage{amsmath,amssymb,graphics}
\usepackage{a4}
\usepackage{epsf}

\numberwithin{equation}{section}

\begin{document}



\newcommand{\MS}{\overline{MS}}
\newcommand{\gMS}{g_{\overline{MS}}}
\newcommand{\LMS}{\Lambda_{\overline{MS}}}
\newcommand{\gs}{g_{\sigma}}
\newcommand{\Ls}{\Lambda_{\sigma}}
\newcommand{\sT}{\frac{\sigma}{T^2}}
\newcommand{\gL}{\widetilde{g}}
\newcommand{\LL}{\widetilde{\Lambda}}
\newcommand{\sig}{\frac{\widetilde{\sigma}}{T^2}}
\newcommand{\hT}{\hat{T}}
\newcommand{\LLAT}{\Lambda_{LAT}}
\newcommand{\gLAT}{g_{LAT}}
\newcommand{\s}{\widetilde{\sigma}}
\newcommand{\fracN}{\frac{1}{N}}

\def\lsim{\raise0.3ex\hbox{$<$\kern-0.75em\raise-1.1ex\hbox{$\sim$}}}
\def\gsim{\raise0.3ex\hbox{$>$\kern-0.75em\raise-1.1ex\hbox{$\sim$}}}


\title{\huge{\bf{Precision Lattice Calculation\\of SU(2) 't Hooft loops}}}
\author{
Philippe de Forcrand\\
\textit{Institut f\"ur Theoretische Physik, ETH H\"onggerberg, CH-8093 Z\"urich, Switzerland}\\
\textit{CERN, Physics Dept., TH Unit, CH-1211 Geneva 23, Switzerland}\\\\
David Noth\\
\textit{Paul Scherrer Institut, CH-5232 Villigen PSI, Switzerland}\\\\}
\maketitle

\abstract{
The [dual] string tension of a spatial 't Hooft loop in the deconfined
phase of Yang-Mills theory can be formulated as the tension of an
interface separating different $Z_N$ deconfined vacua. We review the 1-loop
perturbative calculation of this interface tension in the continuum 
and extend it to the lattice. The lattice corrections are large.
Taking these corrections into account, we compare Monte Carlo measurements
of the dual string tension with perturbation theory, for $SU(2)$.
Agreement is observed at the 2\% level, down to temperatures ${\cal O}(10) T_c$.
}



\section{Introduction}
Because of asymptotic freedom, the running coupling $g(T)$ in an $SU(N)$ Yang-Mills
theory becomes small at high temperature. However, a perturbative calculation of
$SU(N)$ thermodynamic properties faces two obstacles. $g(T)$ only runs logarithmically,
so that subleading orders of the perturbative expansion contribute significantly
as the temperature is lowered towards $T_c$, the confinement/deconfinement
transition temperature. And infrared divergences prevent an analytic, perturbative
treatment altogether beyond some order.
Heroic efforts have been devoted to the calculation of the pressure to this
maximum order $g^6$ \cite{pressure}. The expansion converges poorly.
One may fear that this is a general feature, and that an accurate perturbative 
calculation of any thermodynamic property is only possible at astronomically
high temperatures. The purpose of this paper is to show a counter-example
to this pessimistic view: the spatial 't Hooft loop.

The tension of a spatial 't Hooft loop has been calculated in perturbation theory
to the first three non-trivial orders, and the expansion appears to converge fast
\cite{giovannangeli:03,giovannangeli:04}. We show here that this perturbative calculation agrees with 
non-perturbative Monte Carlo measurements in the $SU(2)$ theory to a high
precision, down to temperatures ${\cal O}(10) T_c$.

The 't Hooft loop has not received the same attention as the Wilson loop
in numerical simulations of $SU(N)$ lattice gauge theories. There are 
several causes for this relative neglect. \\
$\bullet$ First, as 't Hooft has shown \cite{thooft:78}, area law or perimeter law
for Wilson and 't Hooft loops together is forbidden, in the absence of massless
modes. As a result, in $SU(N)$ Yang-Mills theory, an area law is only
observed for the spatial 't Hooft loop (dual to the temporal Wilson loop)
above the deconfinement temperature $T_c$. The associated dual string
tension $\s$ serves as an order parameter for deconfinement
\cite{Korthals-Altes:1999xb,deforcrand:03}. 
The deconfined phase has traditionally received less attention than the 
confined phase, although the situation has been changing with
the experimental search for the quark-gluon plasma. \\
$\bullet$ Second, the intuitive, ``physical'' meaning of a spatial 't Hooft
loop has not been widely appreciated, and it is often considered as an
``exotic'' observable of marginal interest. 
The fact is that a 't Hooft loop enforces a $Z_N$ interface in the Euclidean system,
and that the dual string tension $\s$ is nothing else but the interface tension
(up to a conventional factor $T$).

\begin{figure}[!t]
\centerline{\epsfxsize=3.8cm\epsfbox{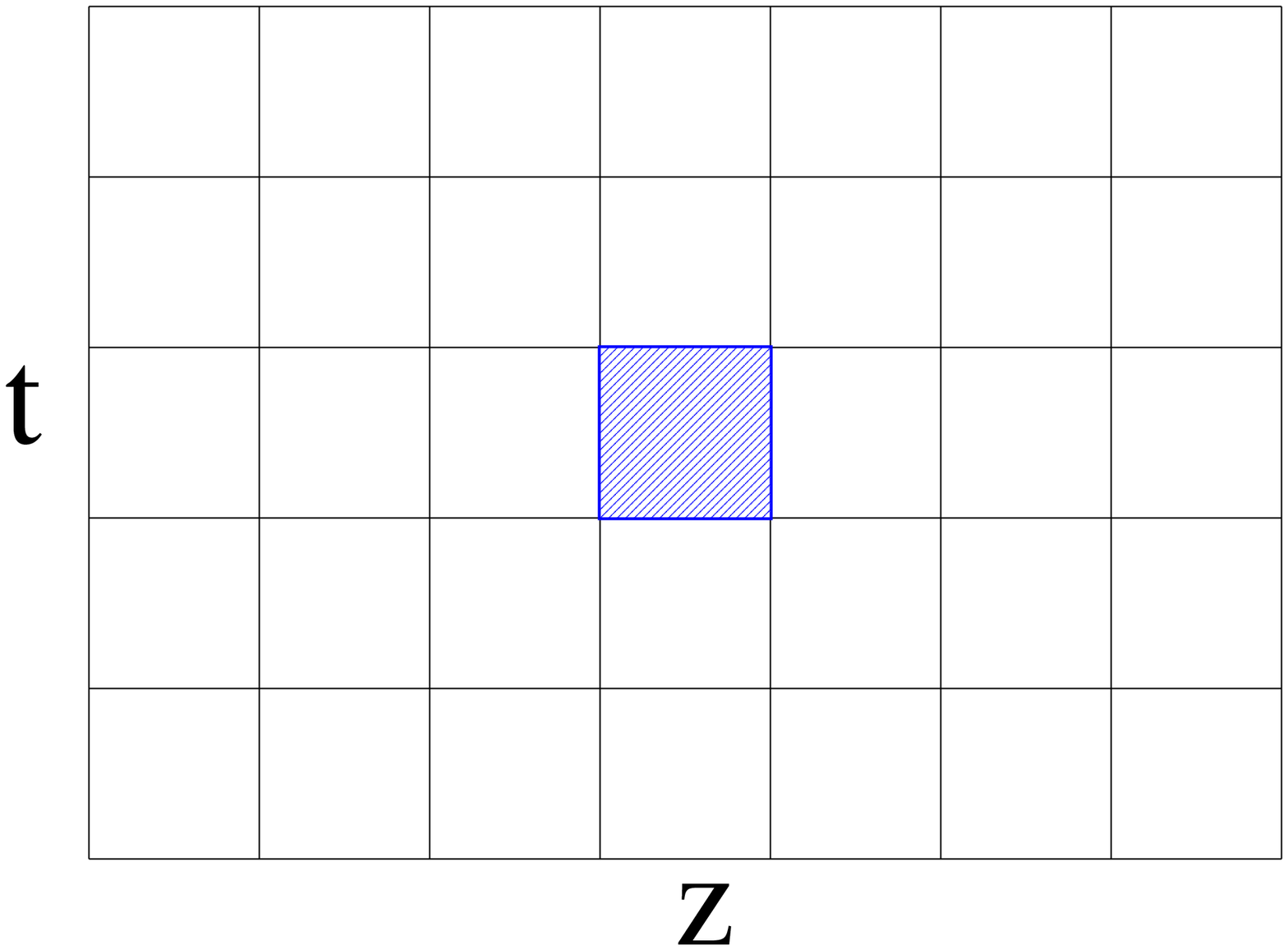}\hspace*{0.8cm}\epsfxsize=3.8cm\epsfbox{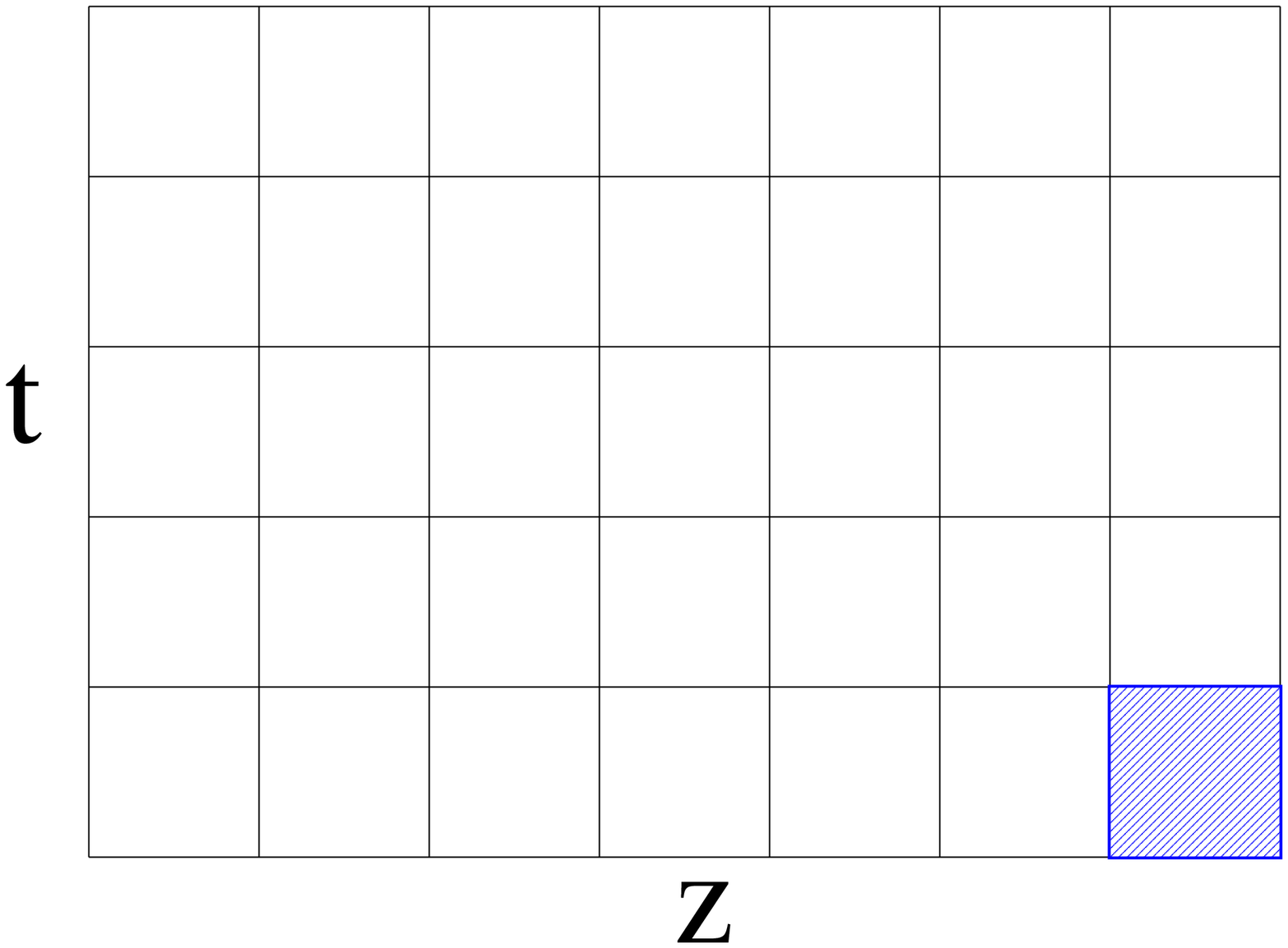}\hspace*{0.8cm}\epsfxsize=3.8cm\epsfbox{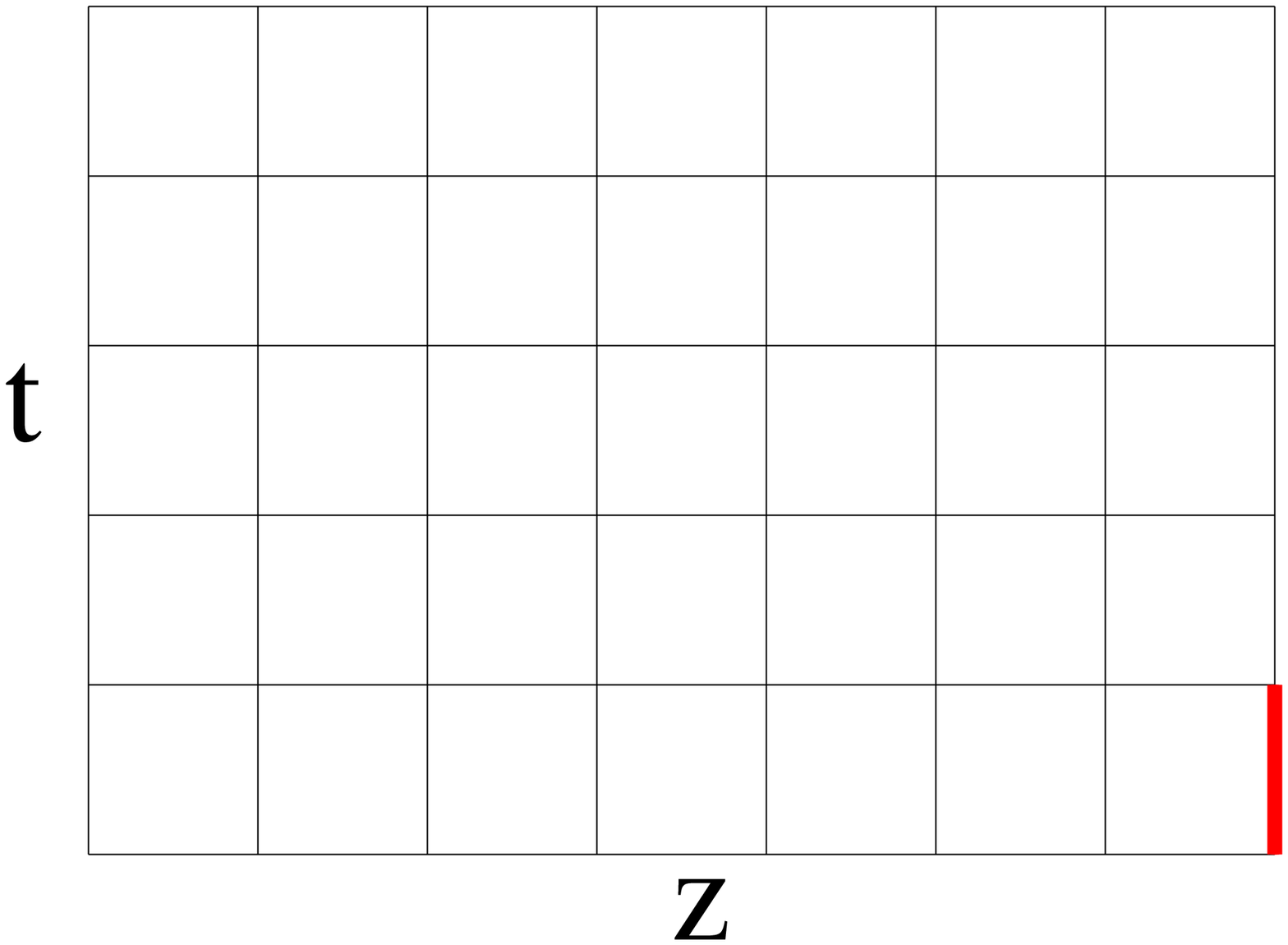}}
\caption{How to create a 't Hooft loop $\partial \tilde\Sigma$ in the $(x,y)$ plane:
{\em (left)} in each $(z,t)$ plane intersecting $\tilde\Sigma$, multiply by a center element one plaquette 
with coordinates $(z_0,t_0)$; {\em (middle)} equivalently, choose $(z_0,t_0)$ in the corner;
{\em (right)} equivalently, multiply by a center element the link $U_t$
at the boundary. Thus, a 't Hooft loop of maximal size ($\tilde\Sigma$ intersects all $(z,t)$ planes) is
equivalent to twisted boundary conditions for the Polyakov loop.}
\label{fig:tbc}
\end{figure}

This is easiest to see on the lattice. Starting from an Euclidean lattice of size
$L_x \times L_y \times L_z \times L_t$, with lattice spacing $a$ ($L_i = N_i a$)
and periodic boundary conditions in all directions, let us construct 
a rectangular 't Hooft loop
$\tilde W(\partial \tilde\Sigma)$ in the $(x,y)$ plane. The contour $\partial \tilde\Sigma$,
which we take rectangular for simplicity, is the boundary of a surface $\tilde\Sigma$,
both on the dual lattice. Let us take for $\tilde\Sigma$ the minimal, planar
surface, with coordinates $(z_0,t_0)$. Each plaquette of $\tilde\Sigma$ is dual
to a $(z,t)$ plaquette. These $(z,t)$ plaquettes with fixed coordinates $(z_0,t_0)$
form a ``stack'' ${\cal P}(\tilde\Sigma)$. The 't Hooft loop expectation value is
then
\begin{equation}
\langle \tilde W(\partial \tilde\Sigma) \rangle \equiv \frac
{\int {\cal D}U \exp(-\beta\sum_{U_P \in {\cal P}(\tilde\Sigma)}(1-\fracN {\rm ReTr}~{\bf \zeta}U_P)
                     -\beta\sum_{U_P \notin {\cal P}(\tilde\Sigma)}(1-\fracN {\rm ReTr}~U_P))}
{\int {\cal D}U \exp(-\beta\sum_{U_P}(1-\fracN {\rm ReTr}~U_P))}
\label{tHdef}
\end{equation}
where $\zeta \in Z_N$ (for $SU(2)$, it amounts to flipping the coupling $\beta \to -\beta$).
Another choice of surface $\tilde\Sigma$ leaves $\langle \tilde W(\partial \tilde\Sigma) \rangle$ invariant,
as can be shown by a change of variables in (\ref{tHdef}): only the contour $\partial \tilde\Sigma$
matters. The dual string tension is then defined by
\begin{equation}
\s \equiv \lim_{A(\partial \tilde\Sigma) \to \infty} -\frac{1}{A} \log \langle \tilde W(\partial \tilde\Sigma) \rangle
\label{sigmadef}
\end{equation}
where $A$ is the minimal area bounded by $\partial \tilde\Sigma$.
We can then consider a 't Hooft loop of maximum size $L_x \times L_y$, equal to that of the lattice in 
the $x$ and $y$ directions. In that case, the stack ${\cal P}(\tilde\Sigma)$ contains 
one plaquette in every $(z,t)$ plane. Following Fig.~\ref{fig:tbc} for each $(z,t)$ plane,
we can move the special plaquette to a corner, then absorb the center element $\zeta$
into the boundary link $U_t$~\footnote{The center element $\zeta$ could be
absorbed in $U_z$ instead, leading to the same free energy for a system with
twisted boundary conditions for the $z$-like ``Polyakov'' loop.}. As a result, the time-like links satisfy
$U_t(x,y,z+L_z,t=1) = \zeta U_t(x,y,z,t=1)$. The same happens for the Polyakov
loop $P(x,y,z) \equiv \prod_t U_t(x,y,z,t)$
(the product of time-like links at position $(x,y,z)$), as
\begin{equation}
\label{Polloop}
        P(x,y,z+L_z) = \zeta P(x,y,z) \quad .
\end{equation}
Therefore, a 't Hooft loop of maximal size in the $(x,y)$ plane is equivalent
to twisted boundary conditions for the Polyakov loop, and
\begin{equation}
\langle \tilde W(L_x, L_y) \rangle = \frac{Z_{tbc}}{Z_{pbc}}
\label{Zratio}
\end{equation}
where the numerator and denominator are the partition functions of a system
with ordinary action, but with boundary conditions respectively twisted (by $\zeta$)
and periodic in the $z$-direction for the Polyakov loop.
Taking the logarithm on both sides, dividing by $L_x \times L_y$, and
taking the thermodynamic limit, one recovers,
on the left-hand side, the dual string tension $\s$, and on the right-hand side,
the interface free energy per unit 2-dimensional area, i.e. the reduced 
interface tension, conventionally written as $\sigma/T$ in a rather confusing
notation. The dual string tension is identical to the reduced interface tension.
Therefore, all the old
studies of $SU(N)$ interface tensions, both numerical \cite{oldSU3} and 
perturbative \cite{bhattacharya:92}, can be re-labeled as 't Hooft loop studies.
It also becomes intuitively clear why $\s$ vanishes below $T_c$:
the Polyakov loop becomes disordered as the $Z_N$ center symmetry is restored,
and the interface tension, i.e. $\s$, vanishes. \\
$\bullet$ Third, the numerical study of the 't Hooft loop has been considered
more difficult than that of the Wilson loop. The reason is an ``overlap''
problem. In the 't Hooft loop expectation value eq.(\ref{Zratio}), the two partition
functions $Z_{tbc}$ and $Z_{pbc}$ are physically different.
Gauge configurations which dominate the integral in the numerator $Z_{tbc}$ 
contain an interface; configurations which dominate in the denominator $Z_{pbc}$
do not. These two ensembles have little overlap, and
importance sampling with respect to the denominator fails.
This technical problem can be approached in various ways, all of which
entail multiple simulations and increased computer cost. Decisive progress
was achieved with the ``snake'' algorithm \cite{deforcrand:03,Pepe} which factorizes the
ratio $Z_{tbc}/Z_{pbc}$ into $N_x \times N_y$ factors, each of ${\cal O}(1)$,
which can each be estimated by an independent Monte Carlo simulation.
In each factor, the area of the interface is increased by one elementary
plaquette $a^2$. Finally, it was realized in \cite{deforcrand:04} that a single
factor converges to $\exp(-\s a^2)$ in the thermodynamic limit.
This brings the computing cost of $\s$ on a par with that of the
string tension $\sigma$ of the Wilson loop (actually, we will see that
the force $\s$ between two dual charges can be computed with a constant
accuracy {\em independent of their separation}, in sharp contrast with the
ordinary string tension). Contact with perturbation
theory for $\s(T)$ becomes feasible at high temperature, as we will show.

One surprise comes on the way. We expect our lattice measurement of
$\s(T)$ to be affected by discretization errors ${\cal O}(a^2)$,
as for any bosonic theory, i.e.
\begin{equation}
        \s(T,a) = \s(T,a=0) (1 + c_1 a^2 + ...)
\end{equation}
where $c_1$ starts with ${\cal O}(g^2)$. Here, $c_1$ starts with 
${\cal O}(g^0)$, which can be traced to the non-perturbative nature of
$\s$. In this paper, we calculate the complete correction
$(\s(T,a)/\s(T,a=0) - 1)$ for $g=0$. It turns out to be
large and quite different from ${\cal O}(a^2)$ for the lattice sizes $N_t$
accessible to current numerical simulations. In fact, it is not even 
monotonic as a function of $N_t$. The expected behaviour 
$\sim c/N_t^2, c \ll 1$, is only recovered for $N_t \gsim 8$. These large 
lattice corrections obscure the analysis of lattice data, and mar the continuum
extrapolations of old $SU(3)$ studies of the critical interface tension
\cite{oldSU3}. 
Our 1-loop calculation of these lattice correction factors,
shown in Table~I and Fig.~\ref{fig:rel_corr}, should be of general use.

Finally, we are in a position to compare our lattice Monte Carlo results for $SU(2)$ with 
continuum perturbation theory \cite{bhattacharya:92,giovannangeli:01,giovannangeli:03,giovannangeli:04}. Remarkable agreement is seen
(cf. Fig.~\ref{fig:GKA}), at the 2\% level, down to temperatures ${\cal O}(10) T_c$,
with perturbation theory at ${\cal O}(g^2)$. 

Our paper is organized as follows. In Sec.~II, we recall for completeness
the perturbative calculation of the interface tension of Ref.~\cite{bhattacharya:92},
then show how this calculation is modified on the lattice,
and extract the lattice correction factors Table I. In Sec.~III, we compare
our Monte Carlo results with the perturbative ones. Conclusions follow.

\section{1-Loop Calculation of the Interface Tension}
\label{sec:1-loop}

\subsection{Continuum Derivation} \label{sec:continuumder}

Following \cite{bhattacharya:92}, we calculate the $Z_2$ interface tension, which is derived from the effective action of a kink interpolating between the two $SU(2)$ vacua. The effective action is the sum of the classical action plus a quantum term, obtained by integrating out fluctuations at 1-loop order.\\\\
The calculation is done in euclidean space-time at temperature $T$. The euclidean time $x_0$ runs from $0$ to $1/T$. The system has spatial size $L_x \times L_y \times L_z$, which is taken to infinity in the end.\\\\
A Z(2) interface along the $z$-direction is constructed by assigning one vacuum to $z=0$ and 
the other to $z=L_z$, and minimizing the effective action subject to these boundary conditions. 
By definition, the interface tension $\s$ is equal to the action of this interface divided by the area of the plane $L_xL_y$ in the thermodynamic limit 
($\s$ is often called the ``reduced'' interface tension; one can equivalently consider the (full) interface tension $\s_F$, which is the interface action divided by the transverse volume $L_xL_y/T$. Of course, $\s = \s_F/T$).\\\\
We consider a gauge field $A$ which is non-trivial only in the $x_0$-direction:
\begin{equation}
        A_{\mu}^{cl}= C_{\mu}\tau_3\ \ ;\ \ C_0=\frac{2\pi T}{g}\:q\ ,\ C_i=0\ \ i=1,2,3
        \label{eq:cd_Acl}
\end{equation}
where $\tau_3$ is the diagonal generator of SU(2) ($\tau_3=\frac{\sigma_3}{2}$). Then the Polyakov loop is
\begin{equation}
        P(\textbf{x}) = \frac12 Tr\left[\mathcal{P}\exp\left(ig\int_0^{\frac1T}dx_0A_0^{cl}(x)\right)\right] = \cos(\pi q)
\end{equation}
The trivial vacuum is at $A_{\mu}^{cl}=q=0$ with $P(\textbf{x})=1$. The $Z(2)$ transform of the trivial vacuum occurs for $q=1$, with
$P(\textbf{x})=\exp(i\pi)=-1$. We can therefore introduce an interface by choosing $q$ in (\ref{eq:cd_Acl}) to be a function of $z$ and fixing $q(0)=0$ and $q(L_z)=1$. Then $P(\textbf{x})=1$ at $z=0$ and $P(\textbf{x})=-1$ at $z=L$.\\\\
We need to justify why the path of minimal action between the two vacua can be brought to
the form (\ref{eq:cd_Acl}), with $q=q(z)$. That is because a constant background field can be brought to the diagonal, 
Cartan subalgebra by a global gauge rotation. But for SU(2), there is only one diagonal generator.\\\\
The \emph{classical action} is
\begin{equation}
        S^{cl}(A)=\int_0^{\frac1T}dx_0\int d^3x\ \frac12 Tr\left[G_{\mu\nu}^2\right]
\end{equation}
with the field-strength tensor $G_{\mu\nu} = \partial_{\mu}A_{\nu}-\partial_{\nu}A_{\mu}-ig[A_{\mu},A_{\nu}]$. For the field in (\ref{eq:cd_Acl}) the classical action is
\begin{equation}
        S^{cl}=\frac{2}{g^2}TL_xL_y\pi^2 \int_0^{L_z} dz \left[\frac{dq}{dz}(z)\right]^2
        \label{eq:cd_Scl}
\end{equation}
Minimizing the action subject to the boundary conditions $q(0)=0,\ q(L_z)=1$, yields $q(z)=z/L_z$. There is no true interface since the action vanishes like $1/L_z$ when $L\rightarrow\infty$ is taken. But this is to be expected: classically there is no energy barrier, since the action is 
constant for any value of $q$.\\\\
This degeneracy of the action with respect to $q$ is broken by quantum effects. To show this, we calculate the action in the presence of the (classical) background field defined in (\ref{eq:cd_Acl}). To do this, we first assume $q$ to be constant in space-time. 
The neglect of gradient terms will be justified a posteriori.\\\\
With $A_{\mu}=A_{\mu}^{cl}+A_{\mu}^{qu}$ the lagrangian consists of the following Yang-Mills term plus gauge fixing term (background field gauge) and Faddeev-Popov ghost term
\begin{eqnarray}
        \mathcal{L} & = & \mathcal{L}_{YM}+\mathcal{L}_{GF}+\mathcal{L}_{FPG} \nonumber\\
        \mathcal{L}_{YM} & = & \frac12 Tr\left[G_{\mu\nu}^2\right] \nonumber\\
        \mathcal{L}_{GF}& = & \frac{1}{\xi}Tr\left[F^2\right] \nonumber\\
        \mathcal{L}_{FPG} & = & Tr\left[ \bar{\eta} P \eta \right]
\end{eqnarray}
where the gauge fixing term is $F=D_{\mu}^{cl}A_{\mu}^{qu}$. The operator $P$ in the ghost part is defined as $P\equiv-2g\frac{\delta_{\Lambda}F}{\delta\Lambda}$, with a gauge transformation $\Lambda$. Also we define the covariant derivative in the adjoint representation for the background field:
\begin{equation}
        D_{\mu}^{cl}=\partial_{\mu}-ig[A_{\mu}^{cl},\ ]
        \label{eq:cd_Dcl}
\end{equation}
The whole lagrangian \emph{quadratic} in the fields is:
\begin{eqnarray}
	\mathcal{L}_{YM}+\mathcal{L}_{GF} & = & Tr\left[A_{\mu}^{qu}(x)\left(-\delta_{\mu\nu}D_{cl}^2
		+(1-\frac{1}{\xi})D_{\mu}^{cl}D_{\nu}^{cl}\right)A_{\nu}^{qu}(y)\right]\delta^4(x-y) \\
	\mathcal{L}_{FPG} & = & Tr\left[ \bar{\eta}(x) \left(-2 D_{cl}^2\right) \eta(y)\right]\delta^4(x-y)
\end{eqnarray}
Using Feynman gauge ($\xi=1$) it follows:
\begin{eqnarray}
	\mathcal{L} & = & \mathcal{L}_{YM}+\mathcal{L}_{GF}+\mathcal{L}_{FPG}\nonumber\\
  & = & \frac12 A_{\mu a}^{qu}(x) \mathcal{M}_{\mu\nu}^{ab}(x,y) A_{\nu b}^{qu}(y)+\bar{\eta}_a(x) \mathcal{F}_{ab}(x,y) \eta_b(y)\\
	\mathcal{F}_{ab}(x,y) & = & \delta^4(x-y)\ \left[-D_{cl}^2\right]_{ab} \label{eq:cd_F} \\
 	\mathcal{M}^{\mu\nu}_{ab}(x,y) & = & \delta_{\mu\nu}\ \delta^4(x-y)\ \left[-D_{cl}^2\right]_{ab} \label{eq:cd_M}
\end{eqnarray}
To obtain the quantum effective action to one-loop $S_1^{qu}$ we have to integrate out the quantum fields $A^{qu},\bar{\eta},\eta$
\begin{eqnarray}
        e^{-S_1^{qu}} & = & \int\mathcal{D}A^{qu}\mathcal{D}\bar{\eta}\mathcal{D}\eta e^{-S}\nonumber\\
        & = & \int\mathcal{D}A^{qu}\mathcal{D}\bar{\eta}\mathcal{D}\eta
                \exp\left[-\frac12\int d^4xd^4y A_{\mu a}^{qu}(x) \mathcal{M}_{\mu\nu}^{ab}(x,y) A_{\nu b}^{qu}(y)\right.\nonumber\\
        & & \left.-\int d^4xd^4y\bar{\eta}_a(x) \mathcal{F}_{ab}(x,y) \eta_b(y)\right]\nonumber\\
        \Rightarrow S_1^{qu} & = & \frac12 Tr\left[\log\left(\mathcal{M}\right)\right] - Tr\left[\log\left(\mathcal{F}\right)\right] \label{eq:cd_S1qu}
\end{eqnarray}
To calculate such traces it is helpful to diagonalize the 'matrices' $\mathcal{M}$ and $\mathcal{F}$ by passing to momentum space.
We then need the eigenvalues of the matrix
\begin{equation}
-\widetilde{D}_{cl}^2(k)=
\left(
        \begin{array}{ccc}
                \mathbf{k}^2+k_0^2+(2\pi Tq)^2 & 2i(2\pi Tq)k_0 & 0\\
                -2i(2\pi Tq)k_0 & \mathbf{k}^2+k_0^2+(2\pi Tq)^2 & 0\\
                0 & 0 & \mathbf{k}^2+k_0^2
        \end{array}
\right)
\label{eq:cd_matrix}
\end{equation}
At finite temperature, $k_0=2\pi Tn$ with $n$ integer, and, using the definition
\begin{equation}
        k_{\pm} \equiv 2\pi T(n\pm q)
        \label{eq:cd_kplus}
\end{equation}
it follows
\begin{equation}
        \begin{array}{lllll}
                \lambda_1 & = & \mathbf{k}^2+[2\pi T(n+q)]^2 & = & \mathbf{k}^2+k_{+}^{2}\\
                \lambda_2 & = & \mathbf{k}^2+[2\pi T(n-q)]^2 & = & \mathbf{k}^2+k_{-}^{2}\\
                \lambda_3 & = & \mathbf{k}^2+k_0^2 & & \rm{(independent\ of\ \ q)}
        \end{array}
        \label{eq:cd_eigenvalues}
\end{equation}
This shift in momenta is a typical effect of a constant background field.\\\\
The sum over $k_+$ is equal to the sum over $k_-$, therefore the quantum action reads,
after substitution of (\ref{eq:cd_F}) and (\ref{eq:cd_M}) into (\ref{eq:cd_S1qu}):
\begin{equation}
        S_1^{qu}=V \sum_{n=-\infty}^{+\infty}\int\frac{d^3\mathbf{k}}{(2\pi)^3}\ 2\ \log(\mathbf{k}^2+k_{+}^{2})
\end{equation}
where terms independent of $q$ (eigenvalue $\lambda_3$) have been dropped.
\\\\Now we need to perform the sums. To do this, we first differentiate with respect to $q$.
\begin{equation}
        \frac{\partial S_1^{qu}}{\partial q} = \frac{TV}{\pi^2} \sum_{n=-\infty}^{+\infty}\int d^3\mathbf{k}\ \frac{k_{+}}{\mathbf{k}^2+k_{+}^{2}}
        = -8T^3V\pi^2\ \sum_{n=-\infty}^{+\infty}(n+q)\ \left|n+q\right|
        \label{eq:cd_infsum}
\end{equation}
The divergent sum is interpreted using zeta-function regularization, with 
$\zeta(s,q) \equiv \sum_{n=0}^{+\infty} (n+q)^{-s}$.
Noting that $\sum_{n=-\infty}^{+\infty}(n+q)\ \left|n+q\right| = \zeta(-2,q) - \zeta(-2,1-q)$,
and using the identity $\zeta(s,q) = \frac{B_{-s+1}(q)}{s-1}$ for $s \le 0$, where $B$ is the
Bernouilli polynomial, one gets $\frac{d S_1^{qu}}{dq} = \frac{4}{3}T^3V\pi^2\ \frac{d}{dq} \left[q^2(1-q)^2\right]$.
Integrating back with respect to $q$, one obtains:
\begin{equation}
        S_1^{qu} = \frac{4}{3}T^3V\pi^2\ \left[q^2(1-q)^2\right]
        \label{eq:cd_S1qubasic}
\end{equation}
\begin{figure}[tb]
        \begin{center}
                \begin{tabular}{ll}
                        \scalebox{0.6}{\includegraphics{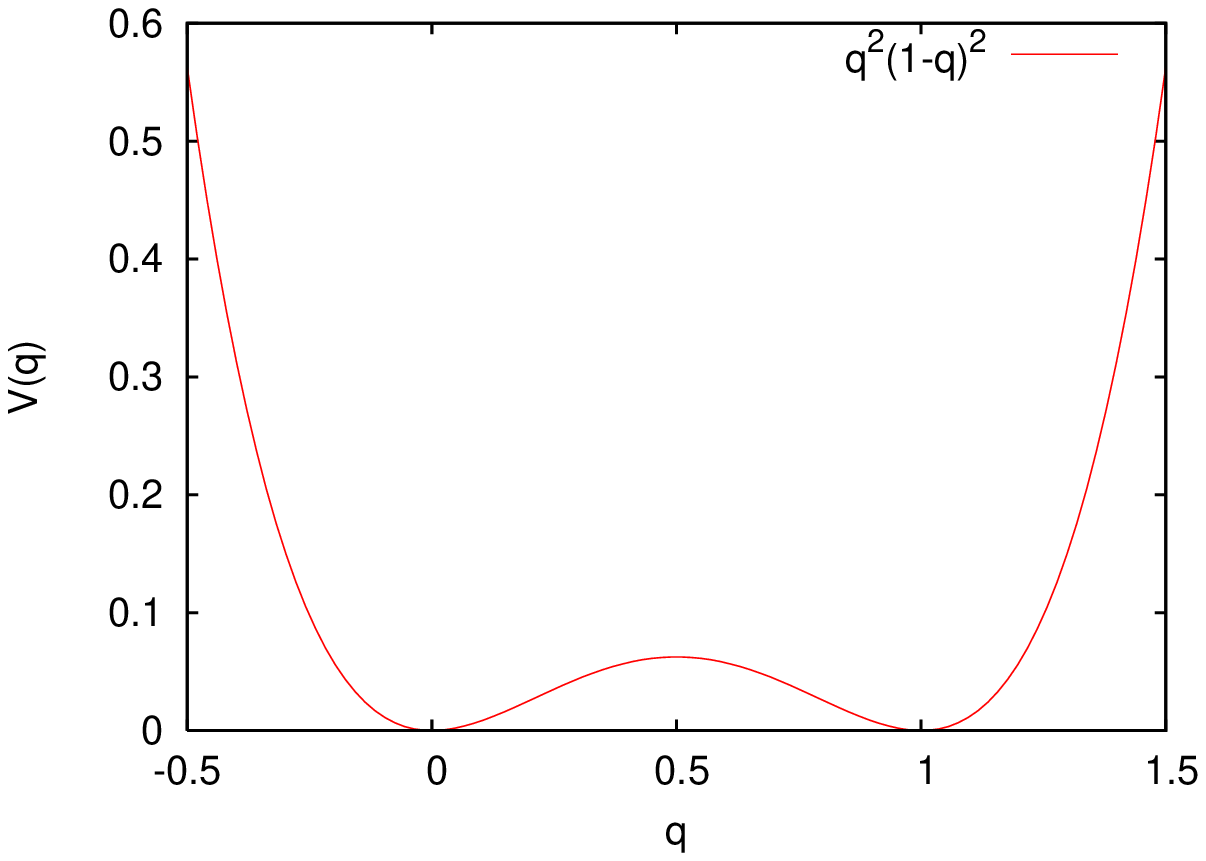}} & \scalebox{0.6}{\includegraphics{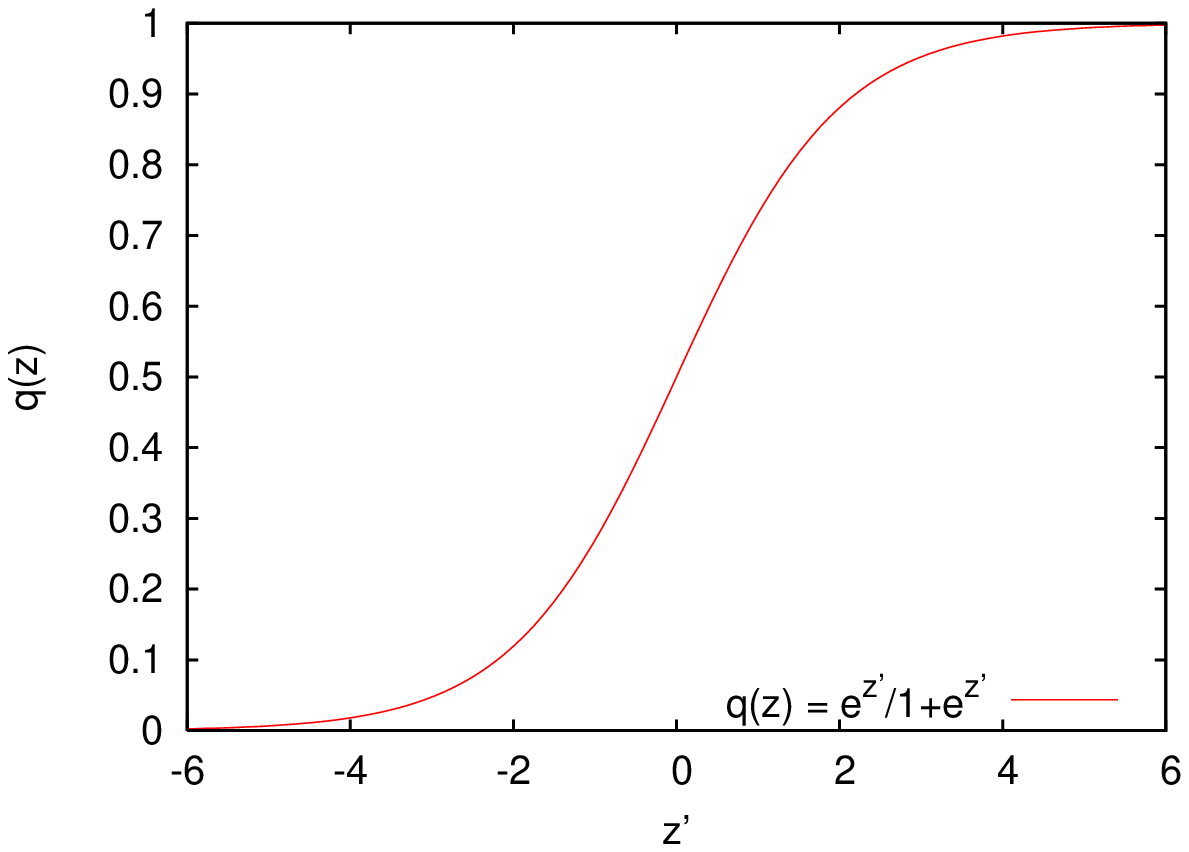}}
                \end{tabular}
        \end{center}
        \caption{Left: The 1-loop effective potential is the function shown in the interval [0,1],
then periodically repeated. 
Right: the kink solution for the effective potential, as a function of $z' = \sqrt{\frac23}gT z$ .}
        \label{fig:cd_Veff}
\end{figure}
So far we treated $q$ as if constant, but actually we want it to be a function $q=q(z)$ of $z$. 
The reason why we can still make use of the action $S_1^{qu}$ for constant $q$ is, that $A^{cl}$ 
varies slowly with $z$, and at leading order this variation can be neglected (gradient expansion),
as we will see in the final kink solution (\ref{eq:kink_sol}). 
Now we reintroduce the $z$-dependence of $q\rightarrow q(z)$ and replace the size of the system 
in the $z$-direction $L_z$ ($V=L_xL_yL_z$) by an integral $L_z\rightarrow\int_0^{L_z} dz$.
\begin{equation}
        S_1^{qu} = \frac{4}{3}T^3L_xL_y\pi^2\ \int_0^{L_z} dz\left[q(z)^2(1-q(z))^2\right]
\label{eq:S1_qu}
\end{equation}
Now the classical degeneracy in $q$ has been lifted by quantum effects and the minima (vacua) are at 
$q=0$ and $q=1$ (see Figure \ref{fig:cd_Veff}, left). Since the sum (\ref{eq:cd_infsum}) 
is periodic in $q$ and hence invariant under shifts $q\rightarrow q+k$ for any integer $k$, 
this is also true for the quantum action. Thus $q$ in $S_1^{qu}$ (\ref{eq:S1_qu}) is defined modulo one.\\\\
From (\ref{eq:cd_Scl}) the classical action is
\[
        S^{cl}=\frac{2}{g^2}TL_xL_y\pi^2 \int_0^{L_z} dz \left[\frac{dq}{dz}(z)\right]^2
\]
Now introduce the dimensionless coordinate $z'=cz$ with $c=\sqrt{\frac23}gT$ in both $S^{cl}$
and $S_1^{qu}$. The complete effective kink action then becomes:
\begin{eqnarray}
        S^{kink}=S^{cl}+S_1^{qu} & = & \sqrt{\frac23}\ \frac{2}{g}\ T^2L_xL_y\pi^2 \int_0^{L'_z} dz'\left[\left(\frac{dq}{dz'}\right)^2+q^2(1-q)^2\right] \label{eq:cd_seffkinpot}
\end{eqnarray}
Defining the Lagrangian $L = \left(\frac{dq}{dz'}\right)^2 + V(q)$, $V(q)\equiv q^2(1-q)^2$,
the Euler-Lagrange equation of motion gives $\frac{dq}{dz'} = \sqrt{V(q)}$.
The solution which satisfies the boundary conditions $q(-\infty)=0$ and $q(+\infty)=1$ is
\begin{equation}
        q(z=z'/c)=\frac{e^{z'}}{1+e^{z'}}
\label{eq:kink_sol}
\end{equation}
This solution is called a \emph{kink} (see Figure \ref{fig:cd_Veff}, right).
The width of the kink is $c^{-1} \sim \frac{1}{g T}$, which is much larger than the scale
$\frac{1}{T}$ of our system since $g \ll 1$. Hence it was indeed justified to neglect
gradient terms $\frac{dq}{dz}$: they only contribute at the next order in $g^2$.
\\\\Thus, we finally have for the effective action of the kink between the two vacua
\begin{equation}
        S^{kink} = \frac{4\pi^2}{3\sqrt6}\ \frac{T^2}{g}L_xL_y
\end{equation}
Our interface tension $\s$ is the kink action per unit planar area, i.e. $S^{kink} / L_xL_y$:
\begin{equation}
        \boxed{\s(T) = \frac{4\pi^2}{3\sqrt6}\ \frac{T^2}{g}}
        \label{eq:cd_interfacetension}
\end{equation}
This formula superficially differs from the corresponding one in \cite{bhattacharya:92} 
by a factor $T$, but this is due to our different definition of the interface tension 
($S^{kink} = \frac{L_xL_y}{T}\s_F$ in \cite{bhattacharya:92}).


\subsection{Corrections on the Lattice} \label{sec:ld_latticecorr}

\begin{figure}[!tb]
        \begin{center}
                \scalebox{0.70}{\includegraphics{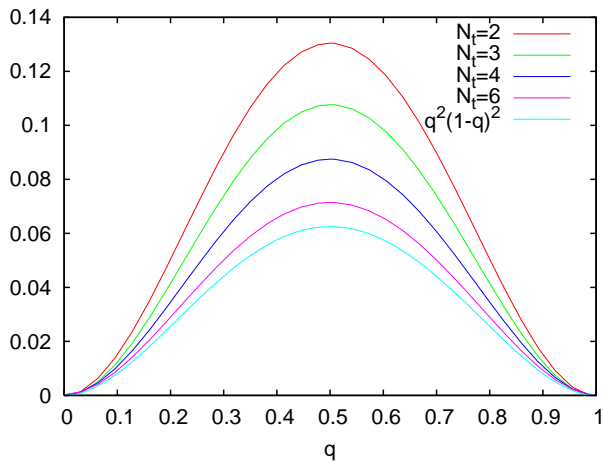}}
        \end{center}
        \caption{The lattice effective potential $V_L(q,N_t)-V_L(0,N_t)$ for different $N_t$'s, 
compared to the continuum effective potential $q^2(1-q)^2$.}
        \label{fig:VLeff}
\end{figure}

\begin{figure}[tb]
        \begin{center}
                \scalebox{0.70}{\includegraphics{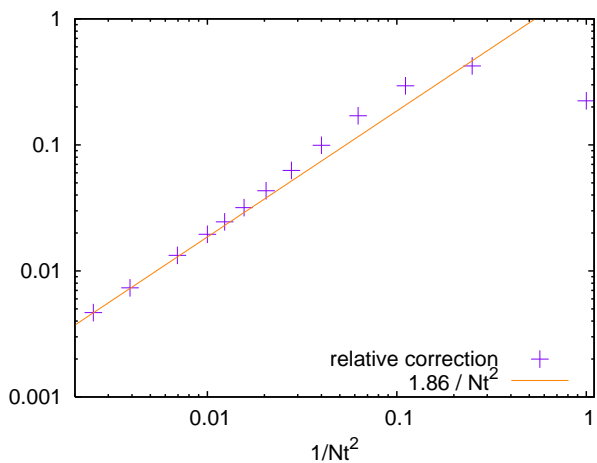}}
        \end{center}
        \caption{The relative correction to the interface tension $\s$, $(C_{lat}(N_t)-1)$, for different values of $N_t$. Only for large values of $N_t$ does the relative correction show the expected 
                behaviour $\propto \frac{1}{N_t^2}$.}
        \label{fig:rel_corr}
\end{figure}

When measuring the interface tension on the lattice, one may expect that the measurements 
of $\sig(N_t)$ ($N_t$ number of temporal sites) approach the perturbative continuum form 
(\ref{eq:cd_interfacetension}) for $g \rightarrow 0$. Somewhat surprisingly, this is not the case.
Multiplicative correction factors are needed, which are large for all $N_t$ accessible to 
numerical simulations. Only for very large $N_t$ do these factors go to $1$. In \cite{weiss:81}, 
Nathan Weiss gave the main expression leading to these correction factors, but without giving 
details about the calculation. This is done here for completeness. Moreover, we show how well 
these results describe the measurements we obtain from Monte Carlo simulations.\\\\
Our derivation on the lattice is very similar to the one in the continuum and in the end, we compare the 
obtained effective potentials from both calculations. Actually the only thing to do is to find 
a lattice expression for the inverse propagator $D_{cl}^{2}$, which took the form (\ref{eq:cd_matrix}) in the continuum.\\\\
A four-dimensional lattice of size $aN_s$ in the spatial directions and $aN_t$ in the temporal direction is used.\\\\
We calculate the classical action first. The Wilson action is
\begin{equation}
        S^{L}_{cl}=\beta\sum_P(1-\frac12 {\rm ReTr}~U_P)
\end{equation}
Only plaquettes in the $zt$-plane are different from $\mathbf{1}$. Such a plaquette at position $z$ takes the value
\begin{equation}
        U_P = e^{ia\frac{\pi}{N_t}\frac1a(q(z+a)-q(z))\sigma_3} = e^{ia\frac{\pi}{N_t}\Delta_z^f q(z)\sigma_3}
\end{equation}
with the forward derivative $\Delta_z^f$. The trace of the plaquette is
\begin{eqnarray}
        2\cos\left(a\frac{\pi}{N_t}\Delta_z^f q(z)\right) & \approx & 2\left( 1 - \frac12 \left(a\frac{\pi}{N_t}\Delta_z^f q(z)\right)^2 \right)
\end{eqnarray}
Placing this into the action we have
\begin{equation}
        S^{L}_{cl} = \beta\sum_{P_{zt}}\ \frac12 \left(a\frac{\pi}{N_t}\Delta_z^f q(z)\right)^2 
                = \frac{2}{g^2}\frac{N_s^2}{N_t}\pi^2 a^2 \sum_z\ \left(\Delta_z^f q(z)\right)^2
\end{equation}
where $\sum_{P_{zt}}$ means sum over plaquettes in the $zt$-plane and $\beta=\frac{4}{g^2}$.\\\\
Mimicking the steps from the continuum derivation, we add quantum fluctuations to this action. 
The calculation on the lattice yields for the momentum space propagator (\ref{eq:cd_matrix})
($k\rightarrow\frac{k'}{a}$ where $k'$ is now a dimensionless quantity)
\begin{equation}
	\left[-\widetilde{D}_{cl}^{2L}\right] = 
  \left(
  \begin{array}{ccc}
	\sum_{\mu}4\sin^2\left(\frac{k'_{\mu}}{2}\right)+4\sin^2\left(\frac{\pi q}{N_t}\right)\cos(k'_0)
		& 2i\sin\left(\frac{2\pi q}{N_t}\right)\ \sin(k'_0) & 0 \\
  -2i\sin\left(\frac{2\pi q}{N_t}\right)\ \sin(k'_0) & \sum_{\mu}4\sin^2\left(\frac{k'_{\mu}}{2}\right)+4\sin^2\left(\frac{\pi q}{N_t}\right)\cos(k'_0)
  	& 0 \\
  0 & 0 & \sum_{\mu}4\sin\left(\frac{k'_{\mu}}{2}\right)
  \end{array}
  \right) \nonumber
\end{equation}
The eigenvalues are
\begin{eqnarray}
        \lambda_1 & = & \frac4{a^2}\left(\sum_{i=1}^3 \sin^2\left(\frac{k_i}{2}\right)+\sin^2\left(\pi\frac{(n+q)}{N_t}\right)\right) \nonumber\\
        \lambda_2 & = & \frac4{a^2}\left(\sum_{i=1}^3 \sin^2\left(\frac{k_i}{2}\right)+\sin^2\left(\pi\frac{(n-q)}{N_t}\right)\right) \nonumber\\
        \lambda_3 & = & \frac4{a^2}\left(\sum_{i=1}^3 \sin^2\left(\frac{k_i}{2}\right)+\sin^2\left(\pi\frac{n}{N_t}\right)\right)
\end{eqnarray}
which can be directly obtained from (\ref{eq:cd_eigenvalues}) by replacing momenta with their lattice counterparts $k \rightarrow \hat{k}=\frac{2}{a}\sin\left(\frac{a}{2} k\right)$.\\\\
The lattice quantum action is (dropping again the eigenvalue $\lambda_3$ since it is independent of $q$):
\begin{eqnarray}
        S_1^{quL} = & V & \sum_{n=0}^{N_t-1}\int_0^{\frac{2\pi}{a}} \frac{d^3\mathbf{k'}}{(2\pi)^3}\ 
                \left\{\log\left[\sum_{i=1}^3 \sin^2\left(\frac{k'_i}{2}\right) + \sin^2\left(\pi\frac{(n+q)}{N_t}\right)\right]\right. \nonumber \\
        & + & \left.\log\left[\sum_{i=1}^3 \sin^2\left(\frac{k'_i}{2}\right) + \sin^2\left(\pi\frac{(n-q)}{N_t}\right)\right]\right\}
\end{eqnarray}
After performing the sum, the lattice quantum action is
\begin{eqnarray}
        S_1^{quL} & = & 2\ N_s^3\ \int_0^{2\pi} \frac{d^3\mathbf{k'}}{(2\pi)^3}\ \log[1-2\cos(2\pi q)e^{-2N_th}+e^{-4N_th}]\\
        h & = & \log(\mathbf{K}+\sqrt{\mathbf{K}^2+1}) \\
        \mathbf{K}^2 & = & \sin^2\left(\frac{k'_x}{2}\right) + \sin^2\left(\frac{k'_y}{2}\right) + \sin^2\left(\frac{k'_z}{2}\right)
\end{eqnarray}
which is the same as stated by Weiss in \cite{weiss:81}.\\\\
We want to compare this to the continuum expression to make a quantitative prediction for corrections to the interface tension measured on the lattice. From (\ref{eq:cd_S1qubasic}) we have in the continuum ($V=L_xL_yL_z$)
\begin{equation}
        S_1^{qu} = \frac{4}{3}VT^3\pi^2\ \left[q^2(1-q)^2\right]
        \label{eq:ld_S1qu}
\end{equation}
Defining
\begin{eqnarray}
        V_L(q,N_t) & = & \left(\frac{3}{2\pi^2}N_t^3\right)\ \int_0^{2\pi} \frac{d^3\mathbf{k'}}{(2\pi)^3}
                \ \log[1-2\cos(2\pi q)e^{-2N_th}+e^{-4N_th}]\nonumber\\
\end{eqnarray}
and subtracting $V_L(q=0)$ from the action in order to make it zero when the system is in one of the vacua we get
\begin{eqnarray}
        S_1^{quL} & = & \frac{4}{3}VT^3\pi^2\ (V_L(q,N_t)-V_L(0,N_t))
\end{eqnarray}
Comparing this to (\ref{eq:ld_S1qu}), we see that $V(q)=q^2(1-q)^2$ is replaced by $V_L(q,N_t)-V_L(0,N_t)$ on the lattice. Both are shown in Figure \ref{fig:VLeff}.
\\\\Taking the continuum limit $N_t \rightarrow \infty$ we have $h \rightarrow \frac{|k|}{2}$ and therefore
\begin{eqnarray}
        V_L(q) & = & \left(\frac{3}{2\pi^2}N_t^3\right)\ \int_0^{\infty} \frac{4\pi k^2 dk}{(2\pi)^3}
                \ \log[1-2\cos(2\pi q)e^{-N_tk}+e^{-2N_tk}]\nonumber\\
        & = & \left(\frac{3}{4\pi^4}\right)\ \int_0^{\infty} x^2 dx
                \ \log[1-2\cos(2\pi q)e^{-x}+e^{-2x}]\nonumber
\end{eqnarray}
which can be shown to be identical to $q^2(1-q)^2\ +\ \rm{constant}$, by differentiating
with respect to $q$.\\\\
This gives for the lattice correction factors $C_{lat}(N_t) \equiv \s(N_t)/\s(N_t=\infty)$:
\begin{equation}
        \boxed{C_{lat}(N_t) = 6\ \int_0^1\ dq \sqrt{V_L(q,N_t)-V_L(0,N_t)}}
\end{equation}
A numerical evaluation yields the factors in Table \ref{tab:ld_corrfactors}.
\begin{table*}
        \small{\begin{center}
                \begin{tabular}[!tb]{|l|l|}       \hline
                        $N_t\ \ \ \ \ \ \ \ \ \ \ $ & $C_{lat}\ \ \ \ \ $                       \\ \hline
          1                                                                                                             & 1.2237                                                        \\ \hline
          2                                                                                                             & 1.4233                                                        \\ \hline
          3                                                                                                             & 1.29467                                                       \\ \hline
          4                                                                                                             & 1.17025                                                       \\ \hline
          5                                                                                                             & 1.09927                                                       \\ \hline
          6                                                                                                             & 1.06278                                                       \\ \hline
          7                                                                                                             & 1.04324                                                       \\ \hline
          8                                                                                                             & 1.03181                                                       \\ \hline
          9                                                                                                             & 1.02451                                                       \\ \hline
          10                                                                                                    & 1.01953                                                       \\ \hline
                \end{tabular}
        \end{center}}
        \caption{Lattice correction factors $C_{lat}$ versus the number $N_t$ of temporal slices.}
        \label{tab:ld_corrfactors}
\end{table*}
As is to be expected in a bosonic theory, the corrections are of order $O(a^2)$, i.e. the factors are of the form $\left(1+O(\frac{1}{N_t^2})\right)$. This is shown in Figure \ref{fig:rel_corr}. 
However, this asymptotic behaviour sets in only for large values of $N_t$. 
For small values of $N_t$, the ones accessible to numerical simulations, the correction factors 
show a quite different, not even monotonic ($N_t=1$), behaviour. This knowledge is crucial when 
trying to extrapolate to the continuum limit with only small-$N_t$ Monte Carlo data.
\\\\To leading order the lattice correction factors $C_{lat}$ do not depend on
the number of colors $N$, which only occurs as a prefactor in the 1-loop
quantum action and cancels out when comparing the continuum to the lattice
effective potential. Thus Table I applies to $SU(N)$ as well.
\\\\There is one more thing to say, concerning finite size effects. So far, we chose the spatial dimensions $N_s$ to be $\infty$. But we can also calculate the impact of a finite size in the 
spatial dimensions on the lattice correction factors. Let us consider a cubic geometry
for simplicity. For finite $N_s$ the integral over the 
momenta $\mathbf{k}$ becomes a sum and we obtain
\begin{eqnarray}
        V_L(q,N_t,N_s) & = & \frac{3}{2\pi^2}\left(\frac{N_t}{N_s}\right)^3\sum_{n_x,n_y,n_z=0}^{N_s-1}
                \ \log[1-2\cos(2\pi q)e^{-2N_th}+e^{-4N_th}] \nonumber\\
        \mathbf{K}^2 & = & \sin^2\left(\frac{\pi}{N_s}n_x\right) + \sin^2\left(\frac{\pi}{N_s}n_y\right) + \sin^2\left(\frac{\pi}{N_s}n_z\right)
\end{eqnarray}
One finds for example at $N_t=2$ and aspect ratio $\rho=2$ 
($\rho \equiv \frac{N_s}{N_t}$) that $C_{lat}$ in Table \ref{tab:ld_corrfactors} should be multiplied 
by 0.885, for $\rho=4$: 0.966 and for $\rho=8$: 0.992. This shows, that with increasing aspect
ratio, the impact of a finite $N_s$ becomes quickly negligible. If finite size
effects are taken into account, $C_{lat}$ decreases slightly, and therefore the corrected
interface tension $\s(T)/C_{lat}$ increases by a tiny amount.

\section{Comparing with Simulations}

\subsection{Measuring the Interface Tension} \label{sec:num_measuringinttension}

A direct Monte Carlo measurement of $\s$ from the free energy definition 
(\ref{Zratio},\ref{sigmadef}) is impractical.
First, there is an overlap problem: configurations contributing to the numerator and
the denominator of (\ref{Zratio}) are physically, macroscopically different, since they respectively
contain an interface or don't. Thus, importance sampling with respect to the
denominator fails for all but the smallest volumes.
Second, the interface is translationally invariant, so that the corresponding
entropy must be carefully subtracted to avoid $\log L$ corrections.
Both problems can be solved by a clever choice of algorithm. Historically,
this choice has been arrived at in two steps.

\begin{figure}[!t]
\centerline{\epsfxsize=8.0cm\epsfbox{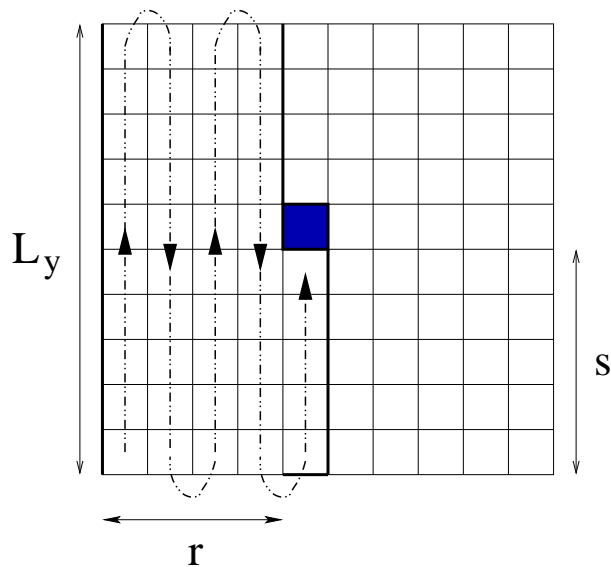}}
\caption{Illustration of the 'snake' algorithm, which builds the interface one plaquette
at a time, and of its simplified version, which requires a single Monte Carlo simulation
of a system with a partial interface.}
\label{fig:snake}
\end{figure}

An elegant approach to the overlap problem is provided by the 'snake' algorithm \cite{deforcrand:03,Pepe}.
It consists of building the interface in $N_x \times N_y$ steps, one plaquette at a time, 
and factorizing the ratio eq.(\ref{Zratio}):
\begin{equation}
\frac{Z_{tbc}}{Z_{pbc}} =
\frac{Z_{N}}{Z_{N-1}}\cdot\frac{Z_{N-1}}{Z_{N-2}}\cdot\dots\cdot\frac{Z_{1}}{Z_{0}}
\label{eq:num_Zratio}
\end{equation}
where $N = N_x \times N_y$, and $Z_k$, $k=0,\dots,N$ ($Z_N\equiv Z_{tbc}$ and $Z_0\equiv Z_{pbc}$) 
is the partition function for a 't Hooft loop of area $k a^2$, with a stack ${\cal P}_k$ of $k$
corresponding plaquettes ``twisted'', i.e. having the coupling $\beta$ changed to $-\beta$:
\begin{equation}
Z_k = \int {\cal D}U \exp(+\beta\sum_{{\cal P}_k}(1-\frac12 {\rm ReTr}~U_P)
                          -\beta\sum_{\bar{{\cal P}_k}}(1-\frac12 {\rm ReTr}~U_P))
\end{equation}
where $\bar{\cal P}_k$ is the complement of ${\cal P}_k$. When $\mod(k,N_y) = 0$, the 't Hooft loop is rectangular
with perimeter $2 N_y$, otherwise it  has perimeter $(2 N_y + 2)$, and 
one side contains 2 kinks (the bold $x$-links in Fig.~\ref{fig:snake}).
The progressive flipping of the plaquettes is described by a snake-like motion, hence the name
attached to the algorithm \cite{Pepe}.
Each ratio in (\ref{eq:num_Zratio}) is of ${\cal O}(1)$ and can be efficiently estimated
by a separate Monte Carlo simulation.

In fact, this algorithm can be simplified \cite{deforcrand:04}, and the $N_x \times N_y$ Monte Carlo simulations
reduced to a single one, by the following observation.
$\frac{Z_{k+1}}{Z_k} = \frac{Z_{k+1}/Z_0}{Z_k/Z_0}$ is a ratio of two 't Hooft loops
differing by $a^2$ in area. If $\mod(k,N_y) \neq \{0,-1\}$, they have the same perimeter and
number of corners. They form the dual analogue of a Creutz ratio, commonly used to measure
the ordinary string tension. Therefore, 
\begin{equation}
\lim_{N_x,N_y \to \infty} -\log\frac{Z_{k+1}}{Z_k} = \s a^2 \quad .
\label{sigma_limit}
\end{equation}
A crucial advantage over the Creutz ratio is that the integrand in $Z_k$ is of exponential
form: it is positive and can be used as a sampling probability. Then, a given statistical
precision on $Z_{k+1}/Z_k$ requires the same computer effort, independently of the 't Hooft
loop size. As stressed in \cite{Caselle:2002ah}, this is in sharp contrast to the ordinary string tension, 
where the statistical accuracy degrades exponentially with the Wilson loop area.
Therefore, (\ref{sigma_limit}) can be rewritten and measured as an expectation value with respect
to $Z_k$:
\begin{equation}
\s a^2 = \lim_{N_x,N_y \to \infty} -\log~\langle \exp(-2\beta \frac12 {\rm ReTr}~U_{P_{k+1}}) \rangle
\label{sigma_limit2}
\end{equation}
In our simulations, we use the equivalent 
$\langle \exp(-\beta \frac12 {\rm ReTr}~U_{P_{k+1}}) \rangle_0 
/ \langle \exp(+\beta \frac12 {\rm ReTr}~U_{P_{k+1}}) \rangle_0$,
where the expectation value $\langle .. \rangle_0$ is taken in the ensemble which interpolates
between $Z_k$ and $Z_{k+1}$, where the coupling attached to $U_{P_{k+1}}$ is set to zero.
This observable has a smaller variance, which can be further reduced by performing multi-hit
(analytically) over the 4 links making up $U_{P_{k+1}}$, since they are now decoupled from
each other.

Note that translation invariance is broken. Our 't Hooft loops $Z_k$ and $Z_{k+1}$ have fixed
boundaries, and the observable is a function of a single plaquette in the system.
This solves the second problem mentioned at the beginning: there is no entropy to
subtract from the free energy of a partial interface.\footnote{In the original 'snake' algorithm,
translation invariance is restored in the last ratio $Z_N/Z_{N-1}$. Its Monte Carlo
evaluation is plagued by very long autocorrelation times, corresponding to translations
of the full interface.} It also suggests to organize the Monte Carlo updates hierarchically
in shells centered around $P_{k+1}$ \cite{Caselle:2002ah}. The gauge links of the innermost shells, upon 
which the observable depends most sensitively, are integrated more thoroughly with more
frequent Monte Carlo updates.

Since taking a large 't Hooft loop in (\ref{sigma_limit2}) requires no more work
than taking a small one, we choose the size as large as practical, as per Fig.~\ref{fig:snake}, with
length $N_y$ and width $r \sim N_x/2$. $r$ can be increased even more, but taking it too close 
to $N_x$ brings about the possibility of reducing the total free energy by trading the
single interface of width $r$ for two of them: one interface of width $(N_x - r)$, and 
one full interface of width $N_x$. Translation invariance of the latter causes the 
free energy reduction. The choice $r \sim N_x/2$ completely eliminates this finite-size effect.

For large 't Hooft loops of size $N_y, r \gg \xi = 1/\sqrt{\s a^2}$, 
the leading deviation of (\ref{sigma_limit2}) from its asymptotic value comes
from two contributions: Gaussian fluctuations of the interface away from its minimal area
(known as capillary waves, or L\"uscher correction), and interaction between the two
kinks in the 't Hooft loop perimeter Fig.~\ref{fig:snake}\footnote{Excited states of the interface
also bring a correction. Assuming an energy gap $\pi/r$ between the ground- and first
excited state propagating along the $y$ direction, one can check that this correction
is smaller than our statistical errors, for all lattice sizes and couplings considered here.}.
$(i)$ The L\"uscher correction can be evaluated analytically. 
For a vibrating $L_y \times r$ interface with $\xi \ll r \ll L_y$, it takes the
familiar form $\s \to \s + \frac{\pi}{12} \frac{1}{r^2}$. In our case where 
$r \sim \frac12 L_y$, it is obtained from the partition function appropriate to
our boundary conditions \cite{Dietz:1982uc}
\begin{equation}
\label{eq:FS_effects_for_sigma} 
Z_{\rm surface}=\eta^2 \left[ i\frac{L_y}{2 r} \right],
\quad \eta(\tau)=e^{i\frac{\pi}{12}\tau}\prod_{n=1}^\infty (1-e^{i 2\pi n \tau}) \ ,
\end{equation}
We subtract this analytic correction from our Monte Carlo results.
$(ii)$ A kink of size $a$ in the 't Hooft loop perimeter creates UV excited string states
of energy ${\cal O}(a^{-1})$, which propagate along the $y$ direction over a distance 
$s \approx a N_y/2$. Therefore, we expect a correction of order $\exp(-N_y/2)$ due to the two kinks.
Indeed, Monte Carlo data obtained on lattices of increasing size $L^3 \times 2$
at fixed $\beta$ are well described by the ansatz 
$(\s a^2 + $ L\"uscher correction $ + c_1 \exp(-c_2 N_y/2))$
with $c_1, c_2 \sim 1$ \cite{deforcrand:04}. This finite-size correction is negligible compared
to our statistical errors provided $N_y \gg -2 \log(\s a^2)$.
Numerical tests at several values of $\beta$ and several lattice extents $N_t$ 
indicated that finite-size effects cause an underestimate of the true
interface tension, and led us to aspect ratios $\{N_x, N_y, N_z\}/ N_t$
as large as 16.

To increase statistics, we typically estimated 2 ratios $Z_{k+1}/Z_k$,
corresponding to plaquettes $U_{P_{k+1}}$ at the ``center'' of the lattice
(i.e. with coordinates $(\frac{N_x}{2},\frac{N_y}{2} + \{0,1\})$).
For each run, an average of 20k multi-hit, multi-shell measurements were
taken.


\subsection{Results versus 2-Loop Perturbation Theory}

\begin{figure}[t]
        \begin{center}
                \begin{tabular}{ll}
                        \scalebox{0.60}{\includegraphics{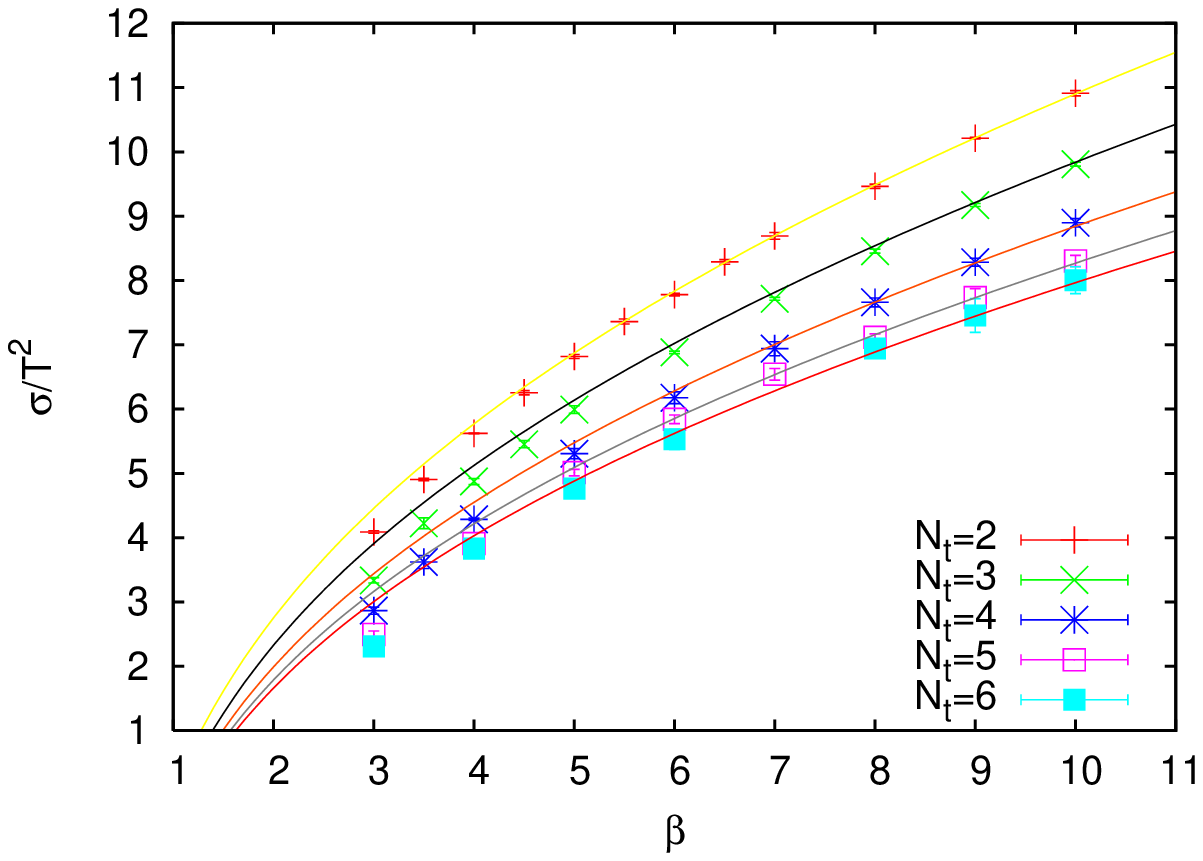}} & \scalebox{0.60}{\includegraphics{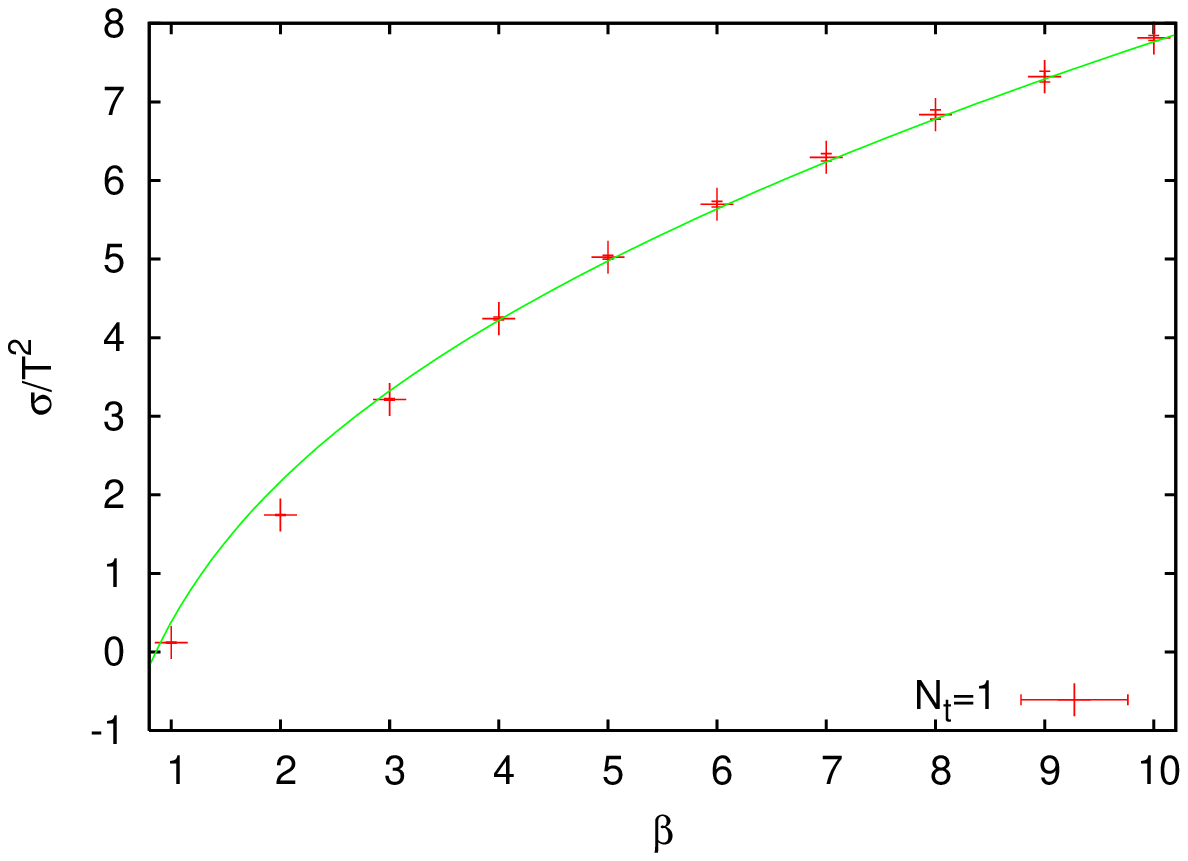}}
                \end{tabular}
        \end{center}
	\caption{The dimensionless ratio of the interface tension over the temperature squared, 
$\sig(N_t,\beta)$, versus $\beta$, for different values of $N_t$. The data at high 
	$\beta$ are described almost perfectly by the {\bf parameter-free} formula (\ref{eq:num_sigmavsbeta}), after the lattice correction factors $C_{lat}(N_t)$ have been inserted.    For $N_t=1$, the perturbative curve describes the measured data almost perfectly over the whole range $\beta>\beta_c$. The reason for this
	remarkably good agreement is, that the singularity in (\ref{eq:num_sigmavsbeta}) occurs for $\beta=0.8587$, which is fortuitously close to the critical
	value $\beta_c=0.8730(2)$ \cite{fingberg:92} of $N_t=1$.}
	\label{fig:sigmaT2_vs_beta}
\end{figure}

In Section \ref{sec:1-loop}, 
we derived the interface tension at 1-loop, in the continuum and on the lattice.
The continuum 2-loop result has been obtained in \cite{bhattacharya:92} and in \cite{giovannangeli:01}, where
the running coupling $\widetilde{g}(T)$ is chosen such that the 1-loop effects disappear in the renormalization of the coupling in the dimensionally reduced theory. For SU(N):
\begin{equation}
        \s(T) = \frac{4\pi^2(N-1)}{3\sqrt{3N}}\ 
                \frac{T^2}{\widetilde{g}(T)}\left(1-\left(15.27853..-\frac{11}{3}(\gamma_E+\frac1{22})\right)\frac{\gL^2(T)N}{(4\pi)^2}\right)
        \label{eq:run_giovanintten}
\end{equation}
with Euler's constant $\gamma_E=0.577215...$. The coupling $\widetilde{g}(T)$ is given in \cite{lissia:01} as
\begin{equation}
        \gL^2(T) \equiv g_{\MS}^2(\mu) \biggm|_{\mu=4\pi T e^{-(\gamma_E+1/22)}}
\end{equation}
which defines the $\LL$-parameter
\begin{equation}
        \LL = \frac{e^{(\gamma_E+1/22)}}{4\pi} \LMS
        \label{eq:run_LLMS}
\end{equation}
With this result it is possible to express the interface tension in
(\ref{eq:run_giovanintten}) to subleading order in terms of the lattice bare
coupling $g_{LAT}(T)$. To do that, we use the $SU(2)$ relation \cite{Hasenfratz:1980kn}
\begin{equation}
  \LMS = 19.8228 \LLAT
\end{equation}
to obtain the 1-loop result
\begin{equation}
  \gL^2 = \gLAT^2(1+0.10016 g^2)
\end{equation}
which is substituted into (\ref{eq:run_giovanintten}) to get
\begin{equation}
        \s(T) = \frac{4\pi^2}{3\sqrt{6}}\ \frac{T^2}{g_{LAT}(T)}\left(1-0.21467 g_{LAT}^2(T) \right)
        \label{eq:run_intten_in_LAT}
\end{equation}
But we want the lattice bare coupling to run with the scale $a^{-1}$ instead of $T=(aN_t)^{-1}$ 
because what is used in a simulation is $\beta=2N/g_{LAT}^2(a^{-1})$. This can be easily 
translated using the $\beta$-function (\ref{eq:new_aMS}). Also the lattice correction factors 
$C_{lat}(N_t)$, calculated in Section \ref{sec:ld_latticecorr} and given in 
Table \ref{tab:ld_corrfactors}, need to be included. Altogether, one gets for \textbf{SU(2)}
\begin{equation}
        \boxed{\sig(\beta, N_t) =
                C_{lat}(N_t)\frac{4\pi^2}{3\sqrt{6}}\ \frac{\sqrt{\beta}}{2} \left(1-(0.21467+\beta_0\log(N_t))\ \frac{4}{\beta} \right)}
        \label{eq:num_sigmavsbeta}
\end{equation}
to subleading order, where $\beta_0$ is given in (\ref{eq:GKA_beta01}).
Note that we have neglected ${\cal O}(g^2)$ corrections in the lattice correction factors
$C_{lat}(N_t)$.
How well this (parameter free!) formula describes the measurements of the interface tension at 
different $N_t$'s for high $\beta$ can be seen in Figure \ref{fig:sigmaT2_vs_beta}. 
The $N_t=1$ comparison is shown separately because the data do not lie neatly above the other 
curves but cross right through.\\\\

\noindent
We have not performed a similar comparison for \textbf{SU(3)}. But analogous steps predict:
\begin{equation}
        \sig(\beta, N_t) =
                C_{lat}(N_t)\frac{8\pi^2}{9}\ \sqrt{\frac{\beta}{6}} \left(1-(0.34805+\beta_0\log(N_t))\ \frac{6}{\beta} \right)
\end{equation}

\subsection{Results versus Temperature}

\begin{figure}[!tb]
        \begin{center}
                \scalebox{0.7}{\includegraphics{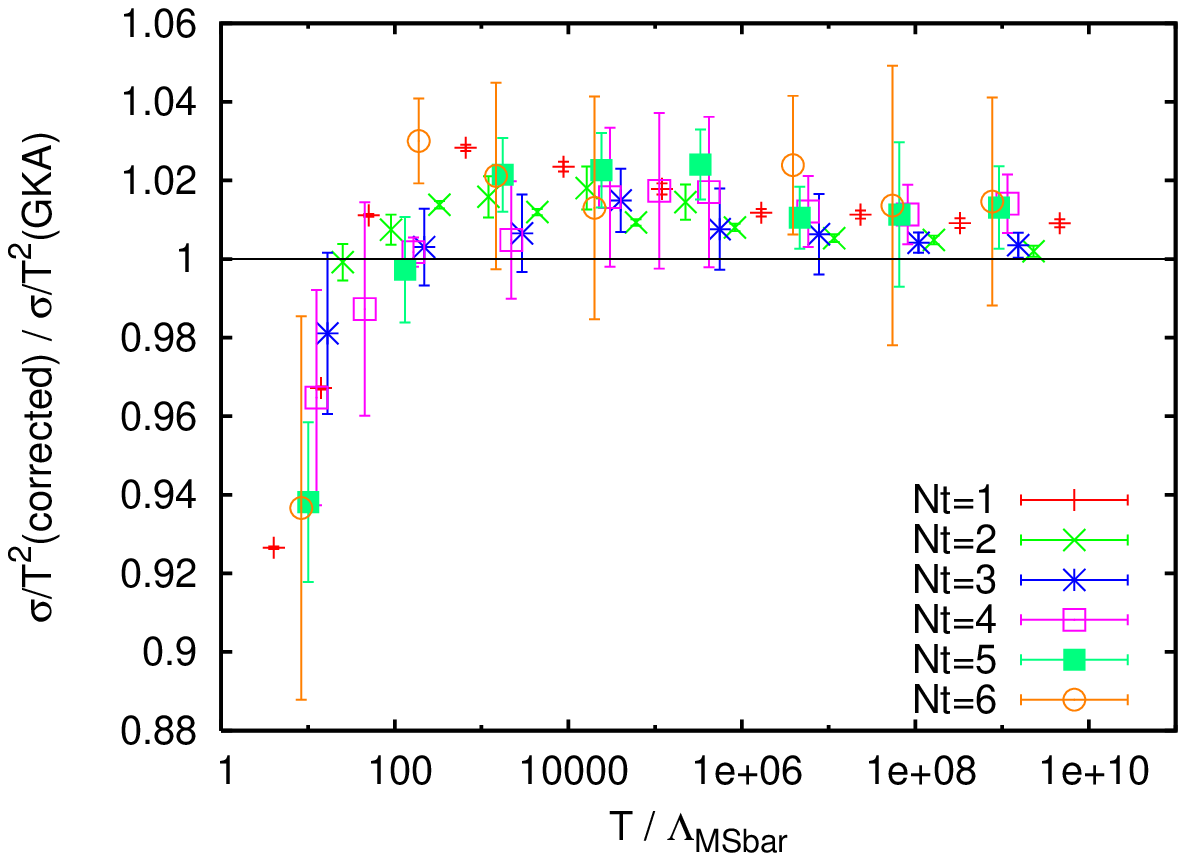}}
        \end{center}
\caption{The ratio of $\sig(\hat{T})_{corrected}$ over $\sig(\hat{T})_{GKA}$ (eq.(\ref{eq:gourGKA})) 
versus $\hat{T}=T/\LMS$. Deviations from $1$ are small ($\sim 2\%$), and become appreciable 
only when $T \lsim 10 T_c$, with $T_c/\LMS = 1.31(8)$ \cite{fingberg:92}.}
\label{fig:GKA}
\end{figure}
Let us collect our measurements of the interface tension for different values of $\beta$ 
and $N_t$, and correct them with the appropriate lattice correction factor $C_{lat}(N_t)$,
forming $\sig(T)_{corrected} = \sig(\beta,N_t)_{measured}/C_{lat}(N_t)$. The bulk of the cutoff effects
has been removed with the division by $C_{lat}(N_t)$, so that the left-hand side can be
compared with the corresponding continuum quantity evaluated in perturbation theory, 
as a function of temperature.
Since we lack the possibility to set a physical temperature scale (to do that, we would need the 
non-perturbative $\beta$-function obtained, e.g., from the step scaling function \cite{luscherwolff:1993}), we calculate 
the temperature $T=1/(a(\beta)N_t)$, associated with each value of $\sig$, 
using the perturbative 2-loop formula for the lattice spacing
\begin{equation}
        a(\beta,\Lambda_{LAT}) = \frac1{\Lambda_{LAT}}
        \exp\left(-\frac{\beta}{8\beta_0}\right)
        \left(\frac{4\beta_0}{\beta}\right)^{-\beta_1/(2\beta_0^2)}
        \label{eq:rvt_a}
\end{equation}
with ($N=2$ here):
\begin{equation}
        \beta_0 = \frac{11}{3}\frac{N}{16\pi^2}\ \ ,\ \ \beta_1 = \frac{34}{3} \left(\frac{N}{16\pi^2}\right)^2
        \label{eq:GKA_beta01}
\end{equation}
which is justified as long as $T \gg \LMS$.
We want to express the temperature in terms of the parameter $\LMS$. One obtains
\begin{equation}
        a(\beta,\LMS) = \frac{19.8228}{\LMS}\exp\left(-\frac{\beta}{8\beta_0}\right) 
                \left(\frac{4\beta_0}{\beta}\right)^{-\beta_1/(2\beta_0^2)}
        \label{eq:new_aMS}
\end{equation}
\noindent
From equation (\ref{eq:run_giovanintten}), we know the relation between $\sig$ and the coupling 
$\widetilde{g}$ at 2-loop order. In \cite{giovannangeli:03}, Giovannangeli and Korthals~Altes show that higher order 
corrections to equation (\ref{eq:run_giovanintten}) are very small. Thus one may hope, that the 
evaluation of (\ref{eq:run_giovanintten}) for $SU(2)$:
\begin{equation}
        \sig(T)_{GKA} = \frac{4\pi^2}{3\sqrt{6}} \frac{1}{\widetilde{g}(T)}\left(1-0.16459\widetilde{g}^2(T)\right)
        \label{eq:gourGKA}
\end{equation}
is sufficient to describe $\sig$ over a large range of temperatures. 
For $\gL^2(T)$ we use the solution to the renormalization group equation to 2-loop: 
\begin{equation}
        \gL^2(T) = \frac1{\beta_0\log(T^2/\LL^2)+(\beta_1/\beta_0)\log\log(T^2/\LL^2)}
        \label{eq:new_gLpert}
\end{equation}
where $\LL$ is related to $\LMS$ through equation (\ref{eq:run_LLMS}).
The perturbative approximation $\sig(T)_{GKA}$ (\ref{eq:gourGKA}) is compared to our measurements
$\sig(T)_{corrected}$ in Figure \ref{fig:GKA}, which shows the ratio $\s(T)_{corrected}/\s(T)_{GKA}$
versus $T/\LMS$. As can be seen, the deviations from $1$ are remarkably small and remain at the
2\% level down to temperatures ${\cal O}(10) T_c$. This figure is a beautiful confirmation of the work of Giovannangeli and Korthals Altes.

\section{Discussion}

We have shown that cutoff effects on lattice measurements of the Yang-Mills
interface tension (or dual string tension) are large for practical lattice sizes
used in Monte Carlo simulations. They do not disappear as $\beta \to \infty$ as 
could be naively expected. Table I, which lists these corrections
calculated at 1-loop order as a function of the temporal lattice size $N_t$
for any $SU(N)$ group, should be of practical use.

Moreover, we have shown precise agreement between our $SU(2)$ Monte Carlo
data and 2-loop continuum perturbation theory, once the correction factors
above have been folded in. The ratio of the measured over the perturbative
interface tension remains 1 within $\sim 2\%$, from very high temperatures
down to ${\cal O}(10) T_c$. Our results put the perturbative calculation
of the interface tension on firm ground, and show that the perturbative
expansion converges well for this quantity, in contrast to the notorious
expansion of the pressure \cite{pressure}.
The reason for this remarkable contrast lies presumably in the topological
nature of the 't Hooft loop. It naturally encodes the $Z_N$ center symmetry,
which is absent in the perturbative expansion of the pressure.

The small remaining discrepancy between the measured data and perturbation theory 
can be assigned various causes. We list them in a subjective order of decreasing
importance:
$(i)$ ${\cal O}(g^3)$ correction in $\s$ \cite{giovannangeli:04}; 
$(ii)$ ${\cal O}(g^2)$ corrections in the lattice correction factors $C_{lat}$;
$(iii)$ 3-loop terms in the lattice $\beta$-function (\ref{eq:rvt_a}),
which will change the $x$-axis $T/\LMS$ in Figure \ref{fig:GKA}; 
$(iv)$ 3-loop terms in the running of $\gL^2(T)$ (\ref{eq:new_gLpert}); 
$(v)$ remaining finite size effects in the Monte Carlo simulations.

Of course, the low-temperature drop in Figure \ref{fig:GKA} is natural.
The interface tension is a (dual) order parameter for the phase transition between the cold 
confining and the hot deconfining phases, which for $SU(2)$ is second-order.
The corresponding correlation length therefore diverges near the critical temperature $T_c$ 
like $(T/T_c - 1)^{-\nu}$, with the critical exponent $\nu$ of the 3d Ising model as expected 
from universality \cite{deforcrand:03}. 
This singularity is not seen by the perturbative calculation.
Hence, the ratio $\s(T)_{corrected}/\s(T)_{GKA}$ in Fig.~\ref{fig:GKA} goes to zero like $(T/T_c - 1)^{2\nu}$.

It would be useful to extend our calculation of the lattice correction factors $C_{lat}$
to order $g^2$, for arbitrary $SU(N)$ group. At this order, one may observe
a dependence on the relative $Z_N$ separation of the two vacua on either side of the interface.
This dependence would enter in the comparison of dual $k$-string tensions
$(\s_k/\s_1)(T)$ between lattice measurements \cite{deforcrand:04} and continuum perturbation theory \cite{giovannangeli:04}.

Finally, we became aware of Ref.\cite{KorthalsAltes:1996xp} as this work was being written up.
Ref.\cite{KorthalsAltes:1996xp} deals with the same issue of connecting lattice measurements and 
continuum calculations of dual tensions, for the $3d$ theory (see Table VIII there). 
Nevertheless, it seems that some of their analytic results generalize to the $4d$ case we studied.

\section*{Note added}
The generalization mentioned above has just been presented in 
Ref.~\cite{Bursa_Teper}. Table I there contains similar lattice correction
factors as our Table I. The small numerical differences can presumably be 
assigned to truncation errors in the series used in Ref.~\cite{Bursa_Teper}.
The numerical simulations in the paper focus on ratios of $k$-interface tensions,
or rather of their derivatives, in $SU(N), N>3$.

\section*{Acknowledgments}

We have benefited from many useful discussions with our colleagues,
including J\"urg Fr\"ohlich, Pierre Giovannangeli, Chris Korthals Altes,
Slavo Kratochvila, Biagio Lucini, Kari Rummukainen and Michele Vettorazzo.
The results presented are part of the work done by D.N. for his diploma thesis
in 2004 at ETH Z\"urich.



\end{document}